\documentstyle[amssymb,aps,epsf]{revtex}

\begin{document}
\title{Space Time Foam: a ground state candidate for Quantum Gravity}
\author{Remo Garattini}
\address{Facolt\`{a} di Ingegneria, Universit\`{a} degli Studi di Bergamo,\\
Viale Marconi, 5, 24044 Dalmine (Bergamo) Italy.\\
E-mail: Garattini@mi.infn.it}
\maketitle

\begin{abstract}
A model of space-time foam, made by $N$ wormholes is considered. The Casimir
energy leading to such a model is computed by means of the phase shift
method which is in agreement with the variational approach used in Refs.\cite
{Remo,Remo1,Remo2,Remo3,Remo4,Remo5}. The collection of Schwarzschild and
Reissner-Nordstr\"{o}m wormholes are separately considered to represent the
foam. The Casimir energy shows that the Reissner-Nordstr\"{o}m wormholes
cannot be used to represent the foam.
\end{abstract}

\section{Introduction}

A very crucial question induced by the appearance of quantum phenomena at
the Planck scale is: what happens when the metric fluctuations become large?
One possible answer should be extracted from the traditional path integral
approach to quantum gravity
\begin{equation}
\int {\cal D}g_{\mu \nu }\exp iS_{g}\left[ g_{\mu \nu }\right] .  \label{p01}
\end{equation}
Unfortunately this quantity is ill defined because the symbol $\left[ {\cal D%
}g_{\mu \nu }\right] $ does not represent a measure. However in the context
of the background field method, a WKB method gives interesting results. In
this context, we can approximate Eq.$\left( \ref{p01}\right) $ with
\begin{equation}
\Gamma =A\exp \left( -I_{cl}\right) ,
\end{equation}
where $A$ is the prefactor coming from the saddle point evaluation and $%
I_{cl}$ is the classical part of the Euclidean action. If a single negative
eigenvalue appears in the prefactor $A,$it means that the related bounce
shifts the energy of the false ground state\cite{Coleman}. In particular in
this approximation, it is possible to discuss decay probabilities from one
space-time to another one \cite{BoHaw,GPY,GP,Young,VW,Prestidge}. For a
certain class of gravitational backgrounds, namely the static spherically
symmetric metrics, it could be interesting the use of other methods based on
variational approach. In a series of papers, we have used such an approach
to show that a model of space-time foam\cite{Wheeler} can be concretely
realized if one considers a collection of Schwarzschild wormholes whose
energy is given by Casimir energy\cite{Remo,Remo1,Remo2,Remo3,Remo4,Remo5}.
We recall that the Casimir energy procedure involves a subtraction procedure
between zero point energies having the same boundary conditions. In this
paper we compare the variational approach with the more traditional phase
shift representation of the Casimir energy. We will consider two types of
static spherically symmetric wormholes: the Schwarzschild wormhole and the
Reissner-Nordstr\"{o}m (RN) wormhole. The final energies will be compared
showing that the Casimir energy for RN wormholes is always higher than the
Casimir energy for the Schwarzschild wormholes. This means that RN wormholes
cannot be taken as a representation of a ground state of a foamy space-time.
To this purpose we will fix our attention on the following quantity
\[
E\left( wormhole\right) =E\left( no-wormhole\right)
\]
\begin{equation}
+\Delta E_{no-wormhole}^{wormhole}{}_{|classical}+\Delta
E_{no-wormhole}^{wormhole}{}_{|1-loop},  \label{p02}
\end{equation}
representing the total energy computed to one-loop in a wormhole background.
$E\left( \text{no wormhole}\right) $ is the reference space energy which, in
the case of the Schwarzschild and RN wormhole, is flat space. $\Delta
E_{no-wormhole}^{wormhole}{}_{|classical}$ is the classical energy
difference between the wormhole and no-wormhole configuration stored in the
boundaries and finally $\Delta E_{no-wormhole}^{wormhole}{}_{|1-loop}$ is
the quantum correction to the classical term. It is possible to proof that
in the considered foam model, the classical term is always vanishing\cite
{Remo,Remo1,Remo2,Remo3,Remo4,Remo5}. The one-loop contribution can be
composed, other than by a stable spectrum, even by an unstable spectrum. If
the unstable spectrum is composed by exactly one element, we can invoke
Coleman arguments to conclude that we move from a false vacuum towards the
true one like in the Euclidean path integral formulation. Nevertheless in
the foam model, we know that the instability can be eliminated in the large $%
N_{w}$-wormhole approach as a wormhole packing consequence\cite
{Remo2,Remo3,Remo4}. Thus to compare two different wormhole models of foam,
it is sufficient to assume the existence of an unstable mode, which will be
subsequently eliminated and compare only the stable spectrum . In
particular, it is the following inequality
\begin{equation}
\Delta E_{flat}^{RN}\left( M,Q\right) \lessgtr \Delta
E_{flat}^{Schwarzschild}\left( M\right)   \label{p02a}
\end{equation}
that will be taken under examination. The final result will give indications
on the possible ``{\it ground state''} of a foamy space-time. Inequality $%
\left( \ref{p02a}\right) $ can be examined by means of Casimir energy, which
in the variational language will be expressed by the following expectation
value\cite{Remo,Remo1,Remo2,Remo3,Remo4}
\begin{equation}
\Delta E_{no-wormhole}^{wormhole}{}_{|1-loop}=\frac{\left\langle \Psi \left|
H_{\Sigma }^{wormhole}-H_{\Sigma }^{no-wormhole}\right| \Psi \right\rangle }{%
\left\langle \Psi |\Psi \right\rangle }.  \label{p03}
\end{equation}
where $\Psi $ is a {\it trial wave functional} of the gaussian form. The
computation of Eq.$\left( \ref{p03}\right) $ can be easily generalized to a
system of $N_{w}$ wormholes such that the hypersurface $\Sigma $ is such
that $\Sigma =\bigcup\limits_{i=1}^{N_{w}}\Sigma _{i}$, with $\Sigma
_{i}\cap \Sigma _{j}=\emptyset $ when $i\neq j$. Thus the total energy will
be simply
\begin{equation}
E^{foam}=N_{w}\Delta E_{no-wormhole}^{wormhole}{}_{|1-loop}.
\end{equation}
This $N_{w}$ wormholes system will be considered as a model for space-time
foam (Units in which $\hbar =c=k=1$ are used throughout the paper).

\section{The wormhole metric and the energy of the foam}

The line element we consider is
\begin{equation}
ds^{2}=-N^{2}\left( r\right) dt^{2}+\frac{dr^{2}}{1-\frac{b\left( r\right) }{%
r}}+r^{2}\left( d\theta ^{2}+\sin ^{2}\theta d\phi ^{2}\right) ,  \label{a1}
\end{equation}
where $N\left( r\right) $ is the lapse function and $b\left( r\right) $ is
the shape function such that\footnote{%
Nothing prevents to consider a positive or negative cosmological constant in
the metric. However, in this paper, the discussion will be restricted to
charged and neutral wormholes.}
\begin{equation}
b\left( r\right) =\left\{
\begin{array}{cc}
2MG & \text{Schwarzschild} \\
2MG-Q^{2}/r & \text{Reissner-Nordstr\"{o}m}
\end{array}
\right. .
\end{equation}
$M$ is the wormhole mass, while $Q^{2}=G\left( Q_{e}^{2}+Q_{m}^{2}\right) $;
$Q_{e}$ and $Q_{m}$ are the electric and magnetic charge respectively. The
wormhole throat $r_{h}$ is located at
\begin{equation}
r_{h}=\left\{
\begin{array}{cc}
2MG & \text{Schwarzschild} \\
\begin{array}{c}
r_{+}=MG+\sqrt{\left( MG\right) ^{2}-Q^{2}} \\
r_{-}=MG-\sqrt{\left( MG\right) ^{2}-Q^{2}}
\end{array}
& \text{Reissner-Nordstr\"{o}m}
\end{array}
\right. .  \label{a1b}
\end{equation}
When $Q=0$ the metric describes the Schwarzschild metric. When $Q=M=0$, the
metric is flat. For $Q\neq 0$, we shall consider only the case $MG>Q$. In a
W.K.B. approximation, Eq.$\left( \ref{p03}\right) $ can be easily computed.
If we restrict to the physical sector of TT (transverse-traceless) tensors,
the Hamiltonian is approximated by
\begin{equation}
H^{\bot }=\frac{1}{4}\int_{\Sigma }d^{3}x\sqrt{g}G^{ijkl}\left[ \left( 16\pi
G\right) K^{-1\bot }\left( x,x\right) _{ijkl}+\frac{1}{16\pi G}\left(
\triangle _{2}\right) _{j}^{a}K^{\bot }\left( x,x\right) _{iakl}\right] ,
\label{p22}
\end{equation}
where we have considered on $\Sigma $ perturbations of the form
\begin{equation}
g_{ij}=\bar{g}_{ij}+h_{ij},
\end{equation}
with $\bar{g}_{ij}$ corresponding to the spatial part of the metric of Eq.$%
\left( \ref{a1}\right) $. The propagator $K^{\bot }\left( x,x\right) _{iakl}$
comes from a functional integration and it can be represented as
\begin{equation}
K^{\bot }\left( \overrightarrow{x},\overrightarrow{y}\right)
_{iakl}:=\sum_{\tau }\frac{h_{ia}^{\left( \tau \right) \bot }\left(
\overrightarrow{x}\right) h_{kl}^{\left( \tau \right) \bot }\left(
\overrightarrow{y}\right) }{2\lambda \left( \tau \right) },
\end{equation}
where $h_{ia}^{\left( \tau \right) \bot }\left( \overrightarrow{x}\right) $
are the eigenfunctions of $\triangle _{2}$. $\tau $ denotes a complete set
of indices and $\lambda \left( \tau \right) $ are a set of variational
parameters to be determined by the minimization of Eq.$\left( \ref{p22}%
\right) $. The expectation value of $H^{\bot }$ is easily obtained by
inserting the form of the propagator into Eq.$\left( \ref{p22}\right) $%
\begin{equation}
E\left( M,Q,\lambda _{i}\right) =\frac{1}{4}\sum_{\tau }\sum_{i=1}^{2}\left[
\left( 16\pi G\right) \lambda _{i}\left( \tau \right) +\frac{E_{i}^{2}\left(
\tau \right) }{\left( 16\pi G\right) \lambda _{i}\left( \tau \right) }\right]
,  \label{p23}
\end{equation}
where we have pointed out the dependence of the energy on some parameters
like the mass and charge. By minimizing with respect to the variational
function $\lambda _{i}\left( \tau \right) $ we get
\begin{equation}
E\left( M,Q\right) =\frac{1}{2}\sum_{\tau }\left[ \sqrt{E_{1}^{2}\left( \tau
\right) }+\sqrt{E_{2}^{2}\left( \tau \right) }\right] .  \label{p24}
\end{equation}
The above expression makes sense only for $E_{i}^{2}\left( \tau \right) >0$,
$i=1,2$. To complete Eq.$\left( \ref{p03}\right) $, we have to subtract the
zero point energy contribution of the space without wormhole: this is the
Casimir energy generated by the curvature potential. In terms of phase
shifts, the Casimir energy is
\begin{equation}
\frac{1}{2}\int_{0}^{+\infty }dpp\sum_{l=0}^{+\infty }\left[ \rho _{l}\left(
p\right) -\rho _{l}^{\left( 0\right) }\left( p\right) \right] ^{\pm }=\frac{1%
}{2\pi }\int_{0}^{+\infty }dpp\sum_{l=0}^{+\infty }\left( 2l+1\right) \frac{%
\partial }{\partial p}\delta _{l}^{\pm }\left( p\right) ,
\end{equation}
where $\rho _{l}\left( p\right) $ represents the density of states in
wormhole background ($\rho _{l}^{\left( 0\right) }\left( p\right) $
represents the density of states in absence of the wormhole, respectively)
and $\delta _{l}^{\pm }\left( p\right) $ is the phase shift due to the
curvature potential. Thus the total Casimir energy is
\[
\Delta E=E^{wormhole}-E^{no-wormhole}
\]
\begin{equation}
=\frac{1}{2\pi }\int_{0}^{+\infty }dp\int_{0}^{+\infty }dl\left( 2l+1\right)
\left[ \left( \frac{d\delta _{l}^{+}\left( p\right) }{dp}+\frac{d\delta
_{l}^{-}\left( p\right) }{dp}\right) \right] p,
\end{equation}
where we have replaced the sum with an integration over all modes. The phase
shift is defined as $\left( r\equiv r\left( x\right) \right) $%
\begin{equation}
\delta _{l}^{\pm }\left( p\right) =\lim_{R\rightarrow +\infty }\left[
\int_{r_{h}}^{x\left( R\right) }dx\sqrt{p^{2}-\frac{l\left( l+1\right) }{%
r^{2}}-\tilde{V}^{\mp }\left( r\right) }-\int_{r_{h}}^{x\left( R\right) }dx%
\sqrt{p^{2}-\frac{l\left( l+1\right) }{r^{2}}}\right] ,
\end{equation}
where $x$ is the proper distance from the throat and $\tilde{V}^{\mp }\left(
r\right) $ is the curvature potential due to the wormhole. If we first
integrate over the angular momenta with the condition that the square root
be real and then we integrate over $p$ with the condition $p\geq \tilde{V}%
^{\mp }\left( r\right) $, we get
\[
\Delta E=\frac{1}{2\pi }\int_{0}^{+\infty }dp\int_{0}^{+\infty }dl\left(
2l+1\right) \left[ \left( \frac{d\delta _{l}^{+}\left( p\right) }{dp}+\frac{%
d\delta _{l}^{-}\left( p\right) }{dp}\right) \right] p
\]
\[
=\frac{1}{\pi }\lim_{R\rightarrow +\infty }\int_{x\left( r_{h}\right)
}^{x\left( R\right) }dxr^{2}\int_{0}^{+\infty }dpp^{2}\left[ \sqrt{p^{2}-%
\tilde{V}^{+}\left( r\right) }+\sqrt{p^{2}-\tilde{V}^{-}\left( r\right) }-2p%
\right]
\]
\[
=\frac{V}{4\pi ^{2}}\left[ \Lambda ^{2}\frac{\tilde{V}^{+}\left(
r_{0}\right) +\tilde{V}^{-}\left( r_{0}\right) }{4}\right.
\]
\begin{equation}
\left. -\left( \frac{\tilde{V}^{+}\left( r_{0}\right) }{4}\right) ^{2}\ln
\left( \frac{\Lambda ^{2}}{\left( \tilde{V}^{+}\left( r_{0}\right) \right)
^{2}}\right) -\left( \frac{\tilde{V}^{-}\left( r_{0}\right) }{4}\right)
^{2}\ln \left( \frac{\Lambda ^{2}}{\left( \tilde{V}^{-}\left( r_{0}\right)
\right) ^{2}}\right) \right] ,  \label{p25}
\end{equation}
where we have introduced a cut-off $\Lambda $ to keep under control the U.V.
divergence, a radius $r_{0}$ such that $r_{0}>r_{h}$ with $r_{0}\neq \alpha
r_{h}$. $\alpha $ is a constant and $V$ is a ``{\it local}'' volume defined
by
\begin{equation}
V=4\pi \int_{x\left( r_{h}\right) }^{x\left( r_{0}\right) }dxr^{2}.
\end{equation}
A comment to justify the approximation leading to Eq.$\left( \ref{p25}%
\right) $ can be useful: since $\tilde{V}^{\mp }\left( r\right) =O\left(
1/r^{3}\right) $, i.e. it is a short range curvature potential and since we
are probing Planckian energies, the contribution to $\Delta E$ comes from
the region close to the throat. One could be tempted to set $r_{0}=r_{h}$.
However a peculiar situation manifests in this limit. Indeed
\begin{equation}
\lim_{M\rightarrow 0}\lim_{r\rightarrow r_{h}}\Delta E\left( M,Q\right) \neq
\lim_{r\rightarrow r_{h}}\lim_{M\rightarrow 0}\Delta E\left( M,Q\right) .
\end{equation}
One possible interpretation of this fact is related to the fluctuation of
the throat enforcing therefore the choice $r_{0}>r_{h}$. The form of $\Delta
E$ changes from a case to case. Here we consider:

\begin{enumerate}
\item  the Schwarzschild wormhole characterized by one parameter: the
wormhole mass $M$.
\begin{equation}
\tilde{V}^{+}\left( r_{0}\right) =\frac{3MG}{r_{0}^{3}},\qquad \tilde{V}%
^{-}\left( r_{0}\right) =-\frac{3MG}{r_{0}^{3}},
\end{equation}
\begin{equation}
\Delta E\equiv \Delta E\left( M\right) =-\frac{V}{32\pi ^{2}}\left[ \left(
\frac{3MG}{r_{0}^{3}}\right) ^{2}\ln \left( \frac{\Lambda ^{2}}{\left(
3MG/r_{0}^{3}\right) }\right) \right] ,
\end{equation}
which is in complete agreement with variational approach used in Refs.\cite
{Remo,Remo1,Remo2,Remo3,Remo4,Remo5}.

\item  The Reissner-Nordstr\"{o}m wormhole characterized by two parameters:
the wormhole mass $M$ and the charge $Q$ with $Q^{2}=G\left(
Q_{e}^{2}+Q_{m}^{2}\right) $; $Q_{e}$ and $Q_{m}$ are the electric and
magnetic charge respectively.

\begin{description}
\item[a)]  electric charge $Q_{e}$%
\[
\tilde{V}^{+}\left( r_{0}\right) =\frac{3MG}{r_{0}^{3}}-\frac{3Q_{e}^{2}}{%
r_{0}^{4}},\qquad \tilde{V}^{-}\left( r_{0}\right) =-\frac{3MG}{r_{0}^{3}}+%
\frac{9Q_{e}^{2}}{r_{0}^{4}},
\]
\[
\Delta E\equiv \Delta E\left( M,Q_{e}\right) =\frac{V}{4\pi ^{2}}\left[
\Lambda ^{2}\frac{3Q_{e}^{2}}{2r_{0}^{4}}\right.
\]
\begin{equation}
\left. -\left( \frac{\tilde{V}^{+}\left( r_{0}\right) }{4}\right) ^{2}\ln
\left( \frac{\Lambda ^{2}}{\tilde{V}^{+}\left( r_{0}\right) }\right) -\left(
\frac{\tilde{V}^{-}\left( r_{0}\right) }{4}\right) ^{2}\ln \left( \frac{%
\Lambda ^{2}}{\tilde{V}^{-}\left( r_{0}\right) }\right) \right] .
\end{equation}

\item[b)]  magnetic charge $Q_{m}$%
\[
\tilde{V}^{+}\left( r_{0}\right) =\frac{3MG}{r_{0}^{3}}+\frac{9Q_{m}^{2}}{%
r_{0}^{4}},\qquad \tilde{V}^{-}\left( r_{0}\right) =-\frac{3MG}{r_{0}^{3}}+%
\frac{Q_{m}^{2}}{r_{0}^{4}},
\]
\[
\Delta E\equiv \Delta E\left( M,Q_{m}\right) =\frac{V}{4\pi ^{2}}\left[
\Lambda ^{2}\frac{5Q_{m}^{2}}{2r_{0}^{4}}\right.
\]
\begin{equation}
\left. -\left( \frac{\tilde{V}^{+}\left( r_{0}\right) }{4}\right) ^{2}\ln
\left( \frac{\Lambda ^{2}}{\tilde{V}^{+}\left( r_{0}\right) }\right) -\left(
\frac{\tilde{V}^{-}\left( r_{0}\right) }{4}\right) ^{2}\ln \left( \frac{%
\Lambda ^{2}}{\tilde{V}^{-}\left( r_{0}\right) }\right) \right] .
\end{equation}
It is immediate to see that the presence of electric and magnetic charge,
respectively gives a positive contribution to the difference of zero point
energies. The final result is displayed in the following plots, where we
have introduced a scale $x$ such that $x=3MG/r_{0}^{3}\Lambda ^{2}$, a
parameter $\alpha _{e}^{2}$ with $0<\alpha _{e}^{2}<1$ for the electric
charge and a parameter $\alpha _{m}^{2}$ with $0<\alpha _{m}^{2}<1$ for the
magnetic charge.
\end{description}
\end{enumerate}

\begin{figure}[tbh]
\vbox{\hfil\epsfxsize=4.5cm\epsfbox{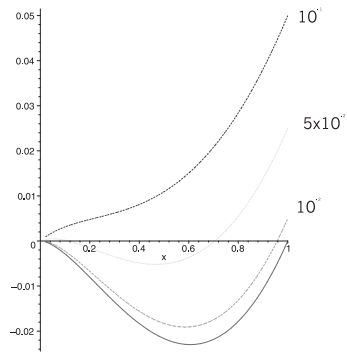}\hfil} \caption{Plot
of Electric charge contribution for different values of the
parameter $\alpha _{e}^{2}$.} \label{f3}
\end{figure}
\begin{figure}[tbh]
\vbox{\hfil\epsfxsize=4.5cm\epsfbox{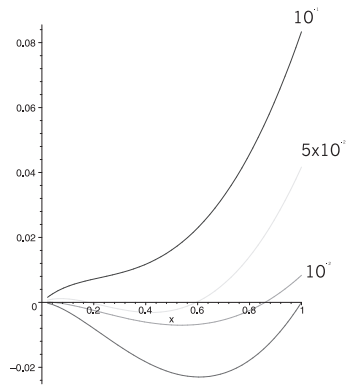}\hfil} \caption{Plot
of Magnetic charge contribution for different values of the
parameter $\alpha _{m}^{2}$.} \label{f4}
\end{figure}
The lowest curve corresponds to the value $\alpha _{e}^{2}=\alpha _{m}^{2}=0$%
, namely the Schwarzschild case.

\section{Summary and Conclusions}

In this paper we have compared the variational approach to compute Casimir
energy with the more traditional phase shift method: the two methods are in
perfect agreement. Moreover we have considered a more general class of
wormholes which includes a charge. This is the RN wormhole class. The
examination of the Casimir energy shows that a space-time foam formation
realized by RN wormholes is suppressed when compared with the foamy space
formed by Schwarzschild wormholes. However, it is an open question the
solution of the problem of a foamy space-time formed by a collection of $N$
extreme RN wormholes. On the other hand one can think to the collection of $%
N $ RN wormholes as an excited state with respect to the collection of $N$
Schwarzschild wormholes leading to the conclusion that such a collection can
be considered as a good candidate for a possible ground state of a quantum
theory of the gravitational field, when compared to a superposition of large
$N$ RN wormholes.

\section{Acknowledgments}

I would like to thank Prof. R. Bonifacio who has given to me the
opportunity of participating to the Conference.

\end{document}